\documentclass[twocolumn,showpacs,floatfix,prb,superscriptaddress]{revtex4}
\usepackage{graphicx}
\usepackage{amsmath}
\usepackage{bm}

\begin{document}
\title{Polarons in suspended carbon nanotubes}
\author{I. Snyman}
\email{isnyman@sun.ac.za}
\affiliation{National Institute for Theoretical Physics, Private Bag X1, 7602 Matieland, South Africa,}
\affiliation{Department of Physics, Stellenbosch University, Private Bag X1, 7602 Matieland, South Africa}
\author{Yu. V. Nazarov}
\affiliation{Kavli Institute of Nanoscience, Delft University of Technology, 2628 CJ Delft, The Netherlands} 
\date{April 2011}
\begin{abstract}
We prove theoretically the possibility of electric-field controlled polaron formation
involving flexural (bending) modes in suspended carbon nanotubes. Upon increasing the field,
the ground state of the system with a single extra electron undergoes a first order phase
transition between an extended state and a localized polaron state. For a common experimental
setup, the threshold electric field is only of order $\simeq 10^{-2}$ V/$\mu$m. 
\end{abstract}
\maketitle
Due to their unique material properies,
carbon nanotubes make ideal flexible nano-rods for mechanical applications \cite{r1}.
Coupling their mechanical motion 
to electronic degrees of freedom leads to non-linear dynamics \cite{r4}.
Current technology \cite{r2} allows for the fabrication of ultra-clean nanotubes
in which electrons propagate ballistically \cite{r1.5} rather than diffusively. 
In combination with a high quality factor \cite{r3}, this allows for resonant excitation and
coherent manipulation of discrete degrees of freedom. 
The envisioned devices may find application in quantum information processing. 

\begin{figure}[tbh]
\begin{center}
\includegraphics[width=.8 \columnwidth]{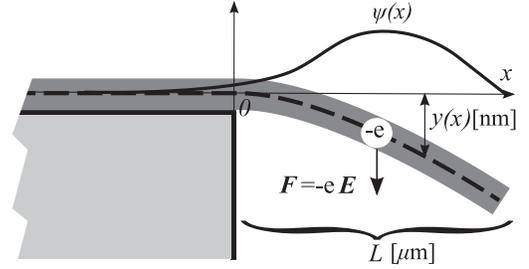}
\caption{Setup: A single wall carbon nanotube cantilever of length $L$.
The supported part of the tube rests on an insulating substrate.
An electron that enters the suspended part of the tube experiences a force $\bm F=-{\rm e}\bm E$
perpendicular to the tube.
As a result the tube is deformed
so that each point $x$ on the tube undergoes a displacement $y(x)$ perpendicular to the $x$-axis.
The electron wave function $\psi(x)$ is also indicated. 
\label{f1}}
\end{center}
\end{figure}

In current devices, a discrete spectrum
is obtained by embedding a quantum dot on a suspended nanotube \cite{r4,r2}.
In this paper we prove the possibility of the controlable formation of discrete states
of a different kind, namely polarons. 
The setup is shown in Figure \ref{f1}.
It consists of an ultra-clean
carbon nanotube cantilever. We consider a single wall semi-conducting nanotube. 
The setup is similar to the nano-relay proposed in Ref. \onlinecite{r5} and to the
experimental setup of Ref. \onlinecite{r6}, but operated in a different
regime, namely that of a single electron on the cantilever. 

If the electron enters the suspended part of the tube,
it experiences a force $\bm{F}=-{\rm e} \bm{E}$. 
The electric field $\bm{E}$ may be due to an external source, or to
an induced image charge in the substrate below the cantilever.
The force $\bm{F}$ deforms the tube.
As a result, the potential energy of the electron is lowered. 
Thus the tube deformation produces a potential well that may trap the electron.

The trapping of an electron in a lattice deformation in a bulk solid is a well-studied topic \cite{r7}.
The resulting quasi-particle is called a polaron.
Previous studies of polarons in carbon nanotubes \cite{r8} considered only 
axial stretching and radial breathing modes of the tube, while our study concentrates on 
macroscopic flexural (bending) modes.

Our main results are contained in Fig. \ref{f2}. At small electric fields, the ground state consists
of an undeformed tube and an extended electron. As the field is increased beyond a critical value,
the system undergoes a first order phase transition  to a localized polaron state. 
For realistic values of a suspended tube length $L=1$ $\mu$m, and tube radius $r=1$ nm, 
the threshold electric field is 
$0.031$ V/$\mu$m and the tip deviation is $0.54$ nm. 
This is the field that would be produced by an image charge induced in
a metallic substrate $0.10$ $\mu$m below the tube.

We start our analysis by noting that 
the typical energy scale for the polaron state is set by the electron confinement energy 
$\varepsilon_e=\hbar^2/2m^*L^2$, where $m^*$ is the effective electron mass.
The ratio $\hbar\omega_0/\varepsilon_e$, where $\omega_0$ is the frequency of the lowest flexural 
mode of the tube, turns out to equal $0.09$, independent of $r$ or $L$.
This ratio is small essentially because electrons weigh much less than carbon atoms.
We therefore neglect the zero point motion (associated with energy $\hbar \omega_0$)
of the cantilever and treat its displacement as a classical variable.
  
The supported part of the tube is tightly clamped to the substrate by van der Waals forces and cannot be deformed.
We take the tube-element that was
at $\bm{r}_0=x\,\hat{\bm{x}}$ 
to be  displaced to $\bm{r}=x\,\hat{\bm{x}}+y(x)\,\hat{\bm{y}}$
under deformation. 
This is valid in the small deflection regime where ${\rm max}\{|y(x)|\}\ll L$.
The system is described by two fields, namely, the tube profile $y(x)$ 
and the wave function $\psi(x)$ of the single electron in the 
conduction band. The boundary conditions on the tube profile are
$y(x\leq0)=y'(x\leq0)=0$ and $y''(L)=y'''(L)=0$.
The boundary conditions on 
the wave function are $\psi(-\infty)=\psi(L)=0$.
The wave function can be taken as real, and is normalized.

The ground state configuration is obtained by minimizing the energy functional
\begin{align}
H[\psi,y]=\int_{-\infty}^Ldx\,\underbrace{\frac{\hbar^2}{2m^*}\left(\partial_x\psi\right)^2}_{=T}
+\underbrace{eEy\psi^2}_{=V}+\underbrace{\frac{YI}{2}\left(\partial_x^2y\right)^2}_{=U}.
\label{a}
\end{align}
The term $T$ is the kinetic energy of the electron. The effective mass
$m^*$ is inversely proportional to the radius $r$ of the nanotube \cite{r9}.
For zig-zag nanotubes
$m^*=1.8\,m_e a_0/r$
where $m_e$ is the true electron mass and $a_0$ is the Bohr radius.
For tubes with chiralities other than zig-zag, the proportionality constant is different, but of the same order of magnitude.  

If the electron
is at position $x$ in the suspended part of the tube, it has undergone a vertical displacement $y(x)$
in the direction of the electrostatic force $-e\bm{E}$. 
This means that the electron sees a potential well with the same profile
as the tube. The term $V$ in Eq.~(\ref{a}) accounts for this.

The term $U$ is the elastic energy of the deformed tube \cite{r10}. In the small deflection approximation,
the energy stored in stretching modes is smaller than the energy stored in flexural modes by a
factor of order $[y(L)/L]^2\ll1$. We therefore only take bending energy into account.
$Y$ is Young's modulus. It is a material constant,
independent of tube dimensions. $I\simeq \pi g r^3$ is the second moment of area
of the tube cross-section. 
Here $g$ is the thickness of the cylinder wall of the nanotube.
Good agreement with nanotube elasticity experiments  
is obtained by taking $g=6.4\,a_0$ (equal to the interlayer distance in graphite) and $Y=1.2\times10^{12}$ Pa \cite{r11}.

It is convenient to introduce dimensionless quantities
$h=\frac{2m^*L^2}{\hbar^2} H$, $\phi=\sqrt{L}\psi$, 
$f=\frac{YI}{eEL^3}y$, and $z=x/L$.
The dimensionless energy functional $h[\phi,f]$ is explicitly given by
\begin{equation}
h[\phi,f]=\int_{-\infty}^1 dz\,(\partial_z\phi)^2+\alpha\left[f\phi^2+\frac{1}{2}(\partial_z^2f)^2\right].\label{h}
\end{equation}
It depends on a single parameter, the dimensionless coupling constant
$\alpha=2m^* (eE)^2 L^5/\hbar^2 YI$.

Two classes of solutions, or phases, can be distinguished in the system. The first class comprises
extended electronic states, in which the magnitude of the wave function is sizable over the whole 
length of the tube. (We consider a tube with total length $\gg L$.)
For such states, the average charge in the suspended
part of the tube is vanishingly small. As a result, the force exerted on the tube by the electric field,
and hence the deformation of the tube, is zero. The total energy of such a
state is equal to the kinetic energy of the electron. Therefore the extended
state spectrum forms a continuum bounded form below by zero. The lowest extended state energy
is zero, corresponding to an electron wave function with an infinite wavelength.  

The second class of states is of the polaron type. These consist of an electron trapped in the potential
well associated with the tube deformation that the electron itself produces. 
The electron wave function $\phi$ decays exponentially into the supported part of the tube, 
i.e $\phi=\phi_0 e^{\kappa z}$
for $z<0$, where $\kappa$ is the inverse localization length. 
Due to the negative potential energy of the trapped electron,
the total energy of the state can become negative. 
When this happens, the ground state of the system is of the polaron variety, since
all extended states have positive energies.
Otherwise the polaron state is meta-stable, since there exists an extended state of zero energy. 
Our task is to determine into which of these two classes the groud state falls for a given value of $\alpha$.
   
In the limit $\kappa\ll1$, where the wave function penetrates
deep into the susported part of the tube, it is straight-forward to estimate the leading order in $\kappa$
contributions of the various terms in Eq.~\ref{h}. (See Appendix~\ref{app1} for detail of the calculation.) 
The energy is dominated by a positive contribution
$\sim\kappa$ of the kinetic energy density in the suspended part of the tube. All other contributions are $\sim\kappa^2$
or smaller. This implies that the transition to the polaron state has to be first order: because 
$\lim_{\kappa\to0^+}\partial_\kappa h$
is positive and cannot change sign, the slope of $h$ can only vanish at  
non-zero $\kappa$.

Since we are dealing with a first, rather than a second order transition, an expansion of the energy in an order 
parameter such as $\kappa$ is of little further use. We therefore proceed to a variational calculation.
For given $\phi$ we find the optimal tube profile $f_0[\phi]$ by minimizing the energy $h[\phi,f]$ with respect to the 
tube profile $f$. This is substituted back into $h$ to obtain $h_{\rm var}[\phi]=h[\phi,f_0[\phi]]$, which is
then varied over a family of trial wave functions $\phi_{\rm var}$. We choose a single parameter family
\begin{equation}
\phi_{\rm var}(z)=N(\kappa)\left\{ \begin{array}{ll}
        e^{\kappa z}& \mbox{if $z<0$;}\\
        \left[(1+\kappa)z+1\right](1-z)&\mbox{if $0\leq z\leq1$.}\label{t}
        \end{array} \right.
\end{equation}
with $N(\kappa)^{-2}=(15+16\kappa+7\kappa^2+\kappa^3)/30\kappa$
ensuring normalization.
The trail wave function has the correct from in the supported part of the tube, is smooth at $z=0$, and
satisfies the boundary condition $\phi(1)=0$ at the suspended end of the tube. The variational parameter
is $\kappa$. Straight-forward but tedious algebra then yields $h_{\rm var}(\kappa)$ as a rational function
where both numerator and denominator are sixth-order polinomials in $\kappa$. (See Appendix \ref{app2} for more detail.)
\begin{figure}
\begin{center}
\includegraphics[width=.8\columnwidth]{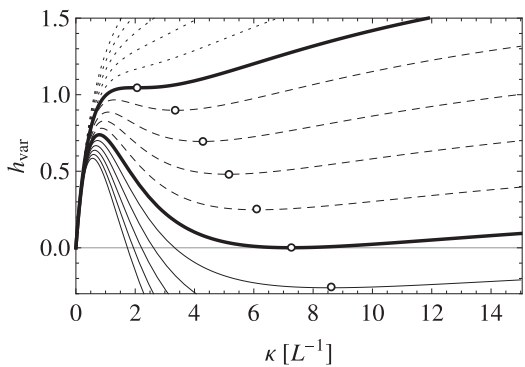}
\caption{First order phase transition: 
The energy $h_{\rm var}$ versus the dimensionless inverse localization length $\kappa$,
calculated for the trail wave function (\ref{t}), for several values of the coupling constant $\alpha$.
The dotted curves correspond to $\alpha<\alpha_{\rm min}^{\rm var}$, where the variational calculation predicts no 
polaron state.
The upper thick curve corresponds to $\alpha=\alpha_{\rm min}^{\rm var}=350.0$.
The dashed curves correspond to $\alpha_{\rm min}^{\rm var}<\alpha<\alpha_{c}^{\rm var}$
where the variational calculation predicts a meta-stable polaron state. 
The thick lower curve corresponds to $\alpha=\alpha_{c}^{\rm var}=431.5$.
The solid curves correspond to $\alpha>\alpha_{c}^{\rm var}$ where the variational curve
predicts a polaron ground state.
Circles indicate the minima corresponding to the polaron state.
\label{f1.5}}
\end{center}
\end{figure}
Note as an aside that, to leading order in $\kappa$, we find $h_{\rm eff}=8\kappa/3$, consistent with the 
discussion in the previous paragraph.
Figure \ref{f1.5} shows $h_{\rm var}$ versus $\kappa$ 
for various values of $\alpha$.

The best estimate for the energy for given $\alpha$ is obtained by 
minimizing $h_{\rm var}(\kappa)$ with respect to $\kappa$ on the interval $[0,\infty)$.
The following results are found: 
For $\alpha<350.0$ (dotted curves in Figure \ref{f1.5}), $h_{\rm var}$ is a monotonically increasing function of $\kappa$ so
that the minimum is $h_{\rm var}=0$ which occurs at $\kappa=0$. 
The implication is that here the ground state is extended.  
At $\alpha>\alpha^{\rm (var)}_{\rm min}=350.0$, a local minimum develops at $\kappa>\kappa_{\rm min}^{\rm var}=2.044$.
For $\alpha^{\rm (var)}_{\rm min}<\alpha<\alpha^{\rm var}_{c}=431.5$ (dashed curves in Figure \ref{f1.5})
this minimum has a positive energy and therefore corresponds to a meta-stable state.
Finally, for $\alpha>\alpha^{\rm (var)}_c$ (thin solid curves in Figure \ref{f1.5}),
the energy of the polaron state becomes negative so that the polaron state is stable.

Next we numerically minimize $h[\phi,f]$ with respect to $\phi$ and $f$, with appropriate
boundary conditions and subject to the constraint that $\phi$ is normalized. Details about the numerical method
can be found in Appendix~\ref{app3}. 

\begin{figure}[tbh]
\begin{center}
\includegraphics[width=.9\columnwidth]{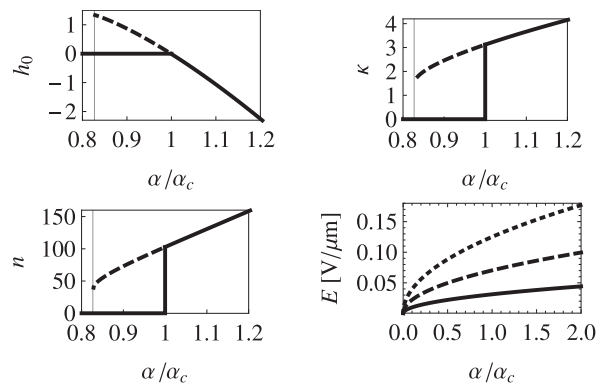}
\caption{Results of the numerical calculation: In the top panels and in the bottom left panel, 
solid curves indicate ground state properties. 
Dashed curves indicate properties of the meta-stable polaron state.
A thin vertical line indicates the value of $\alpha/\alpha_c$ below which no polaron solutions
exist. The critical value of $\alpha$ is $\alpha_c=306.9$. 
Top left: The minimal values $h_0$ of the (dimensionless)
energy $h[\phi,f]$ (cf. Eq. (\ref{h})) vs. $\alpha/\alpha_c$. Top right: The dimensionless inverse localization length
$\kappa$ in the suspended part of the tube vs. $\alpha/\alpha_c$. 
Bottom left: The ratio $n=U/\hbar\omega_0$ between the bending energy and the 
lowest phonon energy vs. $\alpha/\alpha_c$. Bottom right: The electric field $E$ as a function of $\alpha/\alpha_c$,
for a suspended section of length $L=1$ $\mu$m, and three different tube radii $r$. The solid curve is for $r=1$ nm, the 
dashed curve for $r=1.5$ nm and the dotted curve for $r=2$ nm. \label{f2}}
\end{center}
\end{figure}

Thus we find $\alpha_c=306.9$. (See the top-left panel of Figure \ref{f2}.)
This value of $\alpha_c$ is lower than the upper bound derived by means of the
variational calculation above, as it should be. It is also of the same order of magnitude as the 
variational upper bound, indicating that the variational calculation is reasonably accurate. 
We further obtain the value of $\alpha_{\rm min}$, the smallest value of $\alpha$ for which 
polaron states exist as $\alpha_{\rm min}=254.5=0.8293\alpha_c$.

At $\alpha=\alpha_c$ we obtain a critical tip displacement $f_c(1)=0.138$. Reinstating units and eliminating 
the electric field in favour of $\alpha_c$ we obtain $y_c(L)/L=0.138\hbar\sqrt{\alpha_c/2m^*YIL}$.
The critical tip displacement scales like $r^{-1}L^{1/2}$. For realistic values $L=1$ $\mu$m and
$r=1$ nm, we find $y_c(L)=0.54$ nm.

We also calculate $n=U/\hbar \omega_0$, where $U$ is the bending energy  and 
$\omega_0=3.52\sqrt{YI/\rho}/L^2$ \cite{r7} is
the angular frequency of the lowest harmonic of the suspended tube. 
Here $\rho=0.671\,Mr/a_0^2$ is the mass per unit length of the tube, and $M$ is the mass of a carbon atom.
Appendix~\ref{app4} provides more detail.
The quantity $n$, being the ratio between the energy stored in the deformed tube
and the energy of a single phonon, is an estimate of the number of phonons involved in the tube deformation.
In the lower left panel of Figure \ref{f2}, $n$ is plotted as a function of $\alpha$. We see that when the
transition to the polaron state occurs, there are on the order of a hundred phonons in the tube.
The fact that $n$ is large in the polaron state provides additional a posteriori justification for
treating the tube deformation classically.

\begin{figure}[tbh]
\begin{center}
\includegraphics[width=.8\columnwidth]{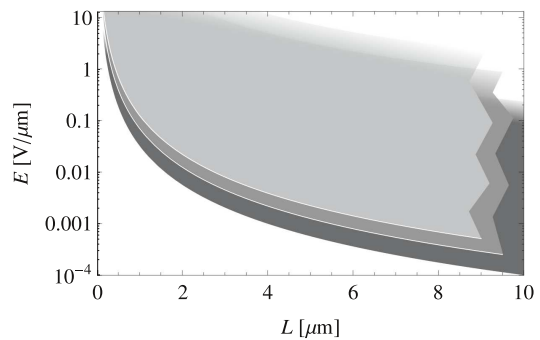}
\caption{Phase diagram: The three different shaded regions correspond to different tube radii (assuming zig-zag tubes):
The dark gray region is for $r=1$ nm, the gray region is for $r=1.5$ nm and the light gray region is for $r=2$ nm.
In each case the shaded region indicates where the ground state is a polaron. 
At large $L$, the upper planes have been cut away to reveal
the planes below. At large $E$ the shaded regions fade to white to emphasize that the upper boundaries
of the shaded regions do not represent a phase transition.\label{f4}}
\end{center}
\end{figure}

An important question to ask is whether values of $\alpha$ that are of the order $\alpha_c$ and larger can
be reached for realistic values of the length $L$, radius $r$, and external electric field $E$. Typical radii are of the
order $1$ nm. Typical lengths are of the order $1$ $\mu$m. An upper bound on the electric field is provided
by the breakdown field of the insulating elements in the setup. These are typically made of ${\rm SiO}_2$ for 
which the breakdown field is $\sim10$ V/$\mu$m. In the bottom right panel of Figure \ref{f2}, we plot the electric
field $E$ versus the corresponding $\alpha$ for $L=1$ $\mu$m and three values of $r$ ranging from $1$ to $2$ nm.
We see that producing a coupling constant in excess of $\alpha_c$ requires an electric field of  
$0.031$ V/$\mu$m for the thinnest tubes and $0.12$ V/$\mu$m for the thickest tubes. 
These are quite reasonable values, well below the breakdown field of ${\rm SiO}_2$.
It is also of the same order as the field produced by an image charge in a metallic substrate $\sim 0.1$ $\mu$m
below the cantelever.

It is informative to draw a phase diagram, indicating the region in parameter space where
the ground state is of the polaron variety. There are two conditions that have to be met. Firstly,
as we have discussed above, the coupling constant must be large enough, i.e. $\alpha(r,L,E)>\alpha_c=306.9$.
There is however another condition: The deflection $y(L)$ of the tube's free tip must be much smaller than
the length $L$ of the suspended section. 
When $|y(l)|\simeq L$, the tube will likely come into contact with one of the surrounding 
elements of the setup, for instance the supporting substrate at $x=0$.
Owing to van der Waals forces the tube will adhere to whatever it comes in contact with.
When this happens, the coupling between electron motion and tube profile is destroyed.
Of course, when this condition is not met, the small deflection approximation, 
on which our analysis relied, also breaks down. Appendix~\ref{app5} provides more detail. 
In Figure \ref{f4} we show three cuts through the phase diagram in the $E$--$L$ plane,
for $r=1,\,1.5,$ and $2$ nm respectively. In each case the polaron phase is indicated by a shaded region. We see that
the value of the largest allowed electric field is always several orders of
magnitude larger than the smallest allowed electric field. 

In conclusion, we found that the coupling between the electron and the
tube is controlled by a dimensionless coupling constant $\alpha=2m^* (eE)^2 L^5/\hbar^2 YI$.
At strong coupling (large $\alpha$) the ground state of the system is a polaron, i.e. the electron is
trapped in the deformation of the cantilever that it itself produces. As the coupling is 
decreased beyond the critical value of $\alpha_c=306.9$, a first order phase transition occurs. Below the transition
point, the ground state consists of an undeformed tube and an extended electron wave function.   
For realistic values $L=1$ $\mu$m for the length of the suspended tube and $r=1$ nm for the tube
radius, an electric field of $0.031$ V/$\mu$m is required to realize the polaron phase and
the critical tip displacement is $0.54$ nm.
The magnitude of the threshold electric field is the same as that produced by
an image charge in a metallic substrate $0.10$ $\mu$m below the cantelever.

In future work we plan to study the nonlinear dynamics of a single polaron as well as the 
interaction between polarons in the same or adjacent suspended tubes. The eventual aim is
to exploit the coupling between mechanical and electrical degrees of freedom for
the coherent manipulation of the quantum state of the polaron.

\appendix

\section{Expanding the energy in small $\kappa$}
\label{app1} 
In the main text it is stated that 
at small inverse localization lengths $\kappa$, the energy $h$ is dominated by a
positive contribution of order $\kappa$. Here we provide a detailed analysis.

Setting 
\begin{equation}
\frac{\delta}{\delta\phi(x)}(h-\varepsilon\int_{-\infty}^1dz\,\phi^2)=\frac{\delta}{\delta f(x)}
(h-\varepsilon\int_{-\infty}^1dz\,\phi^2)=0
\end{equation} 
and using the boundary conditions 
\begin{subequations}
\begin{eqnarray}
&&\phi(z=1)=\lim_{z\to-\infty}\phi(z)=0,\label{k}\\
&&f(z=0)=\partial_zf(z=0)=0,\label{kk}\\
&&\partial^2_zf(z=1)=\partial^3_zf(z=1)=0,\label{l}
\end{eqnarray}
\end{subequations}
we obtain
two differential equations
\begin{subequations}
\begin{eqnarray}
\varepsilon\phi(z)&=&-\partial_z^2\phi(z)+\alpha f(z)\phi(z),\label{m}\\
\partial_z^4f(z)&=&-\phi(z)^2.\label{n}
\end{eqnarray}
\end{subequations}
Here $\varepsilon$ is a Lagrange multiplier that enforces the normalization of $\phi$.
The first of these equations is the Schr\"odinger equation for the electron in a potential 
$\alpha f(z)$. The second equation describes the balance of the electrostatic force 
that deforms the tube and the elastic restoring force. 

The solution to Eq.~(\ref{n}) that satisfies the boundary conditions (\ref{kk}) and (\ref{l}) is
\begin{align}
f_\phi(z)=-\frac{1}{6}&\left[\int_0^zdz'\,(z')^2(3z-z')\phi(z')^2\right.\nonumber\\
&+\left.\int_z^1dz'\,z^2(3z'-z)\phi(z')^2\right].\label{o}
\end{align}
We can substitute this solution into $h$ in order to obtain an effective energy functional 
that depends on $\phi$ only, i.e. $h_{\rm eff}[\phi]=h[\phi,f_\phi]$. By exploiting the fact that
\begin{equation}
\int_0^1dz'\,(\partial_z^2f_\phi)^2=\int_0^1dz'\,f_\phi\partial_z^4f_\phi=-\int_0^1dz'f_\phi\phi^2,\label{p}
\end{equation}
we obtain 
\begin{equation}
h_{\rm eff}[\phi]=\int_{-\infty}^1 dz\,(\partial_z\phi)^2+\frac{\alpha}{2}\int_{0}^1 dz\,f_\phi\phi^2.\label{q}
\end{equation}

We now expand $h_{\rm eff}$ in the inverse localization length
$\kappa$ of the polaron state. Let us firstly look at the kinetic energy. We consider separately the
contribution of the wave function in the supported and suspended parts of the tube.
As mentioned before, the wave function is of the form $\phi=\phi_0 e^{\kappa z}$ in the supported part of the 
nanotube.
In the limit $\kappa\ll1$, the normalization constant $\phi_0$ is of the order $\sqrt{\kappa}$.
For the kinetic energy density integrated over the supported part of the tube, we obtain
\begin{equation}
\int_{-\infty}^0dz\,(\partial_z\phi)^2=\kappa^2\int_{-\infty}^0dz\,\phi^2\sim\kappa^2.\label{r}
\end{equation}
(In the last step we approximated $\int_{-\infty}^0dz\,\phi^2\simeq1$, neglecting the possibility to find
the electron in the region $0<z<1$, which is valid for $\kappa\ll1$.)
To estimate the kinetic energy stored in the suspended part of the tube, we note that
here $\phi$ changes from $\phi_0\sim\sqrt{\kappa}$ at $z=0$ to $\phi=0$ at $z=1$. This corresponds
to a typical slope $\partial_z\phi\sim-\sqrt{\kappa}$ so that 
\begin{equation}
\int_0^1 dz\,(\partial_z\phi)^2\sim\kappa\label{s}
\end{equation}
Thus, at small $\kappa$, the kinetic energy is dominated by a contribution of order $\kappa$. 
Note also that the kinetic energy is positive.
To estimate the remaining term (=$\alpha\int_0^1dz\,f_\phi\phi^2$) in $h_{\rm eff}$
we note from that Eq. (\ref{o}) that $f_\phi \phi^2$ is quartic in $\phi$.
The integral therefore scales like $\phi_0^4\sim\kappa^2$. It is also negative.

\section{Variational calculation}
\label{app2} 
In the main text we discuss a variational calculation in
order to obtain an estimate $h_{\rm var}(\kappa)$ of the energy of the polaron state,
that depends on a single variational parameter $\kappa$. In Fig. 2 of the main text, 
$h_{\rm var}(\kappa)$ is plotted for several values of the coupling constant $\alpha$.
Here we give the explicit formula for $h_{\rm var}(\kappa)$. It is a rational function 
\begin{equation}
h_{\rm var}(\kappa)=\frac{\sum_{n=0}^6 (a_n-b_n\alpha)\kappa^n}{\sum_{n=0}^6 c_n\kappa^n},
\end{equation}
with coefficients
\begin{center}
\begin{tabular}{|r|r|r|r|}
\hline
\multicolumn{1}{|c|}{$n$} & \multicolumn{1}{|c|}{$a_n$} & \multicolumn{1}{|c|}{$b_n$} & \multicolumn{1}{|c|}{$c_n$}\\
\hline
0& 0& 0& 225\\
1& 600& 0& 480\\
2& 1165& 1.639& 466\\
3& 990& 2.151& 254\\
4& 445& 1.087& 81\\
5& 105& 0.2498&14\\
6& 10& 0.02205& 1\\
\hline
\end{tabular}
\end{center} 

\section{Numerical calculation}
\label{app3}

In the main text we present results of a numerical minimization of the energy $h$. Here we provide
some details about the numerical method.
\begin{figure}[tbh]
\begin{center}
\includegraphics[width=\columnwidth]{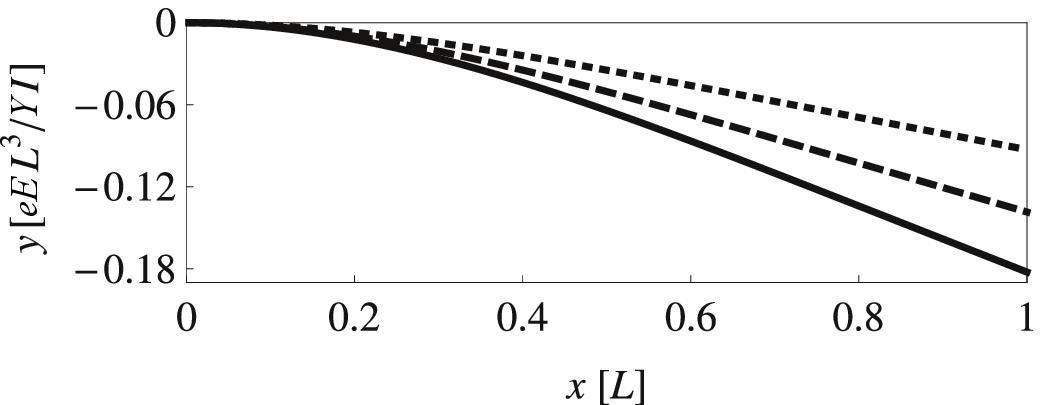}
\includegraphics[width=\columnwidth]{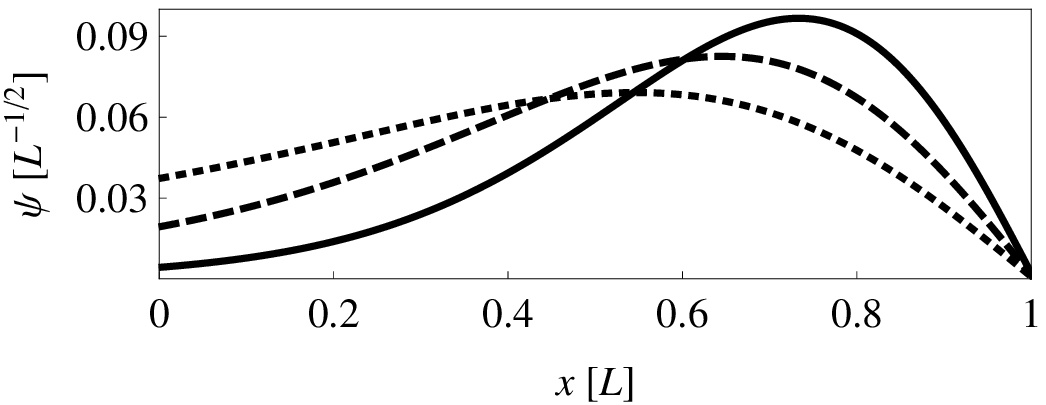}
\caption{Numerically computed polaron states: 
Tube profiles $y(x)$ and wave functions $\psi(x)$ that solve Eqs. (\ref{m}) and (\ref{n}) 
for three different values of the coupling constant $\alpha$. The solid curves are for $\alpha=500=1.629\alpha_c$.
Since $\alpha>\alpha_c$ for the solid curves, they represent the ground state of the system at this particular value of $\alpha$.
The dashed curves are for $\alpha=306.9=1.00\alpha_c$.
Since $\alpha=\alpha_c$ for these curves, they occur at the boundary between stability, where the
energy of the polaron state is negative, and meta-stability where the energy of the polaron state is positive.
The dotted curves are for $\alpha=254.5=0.8293\alpha_c=\alpha_{\rm min}$.
There are no polaron states for $\alpha<\alpha_{\rm min}$.        
\label{f3}}
\end{center}
\end{figure}

We extremize the energy functional $h$ by solving solving 
Eqs.~(\ref{m}) and (\ref{n}) numerically, subject to the boundary conditions (\ref{k}), (\ref{kk}) and (\ref{l}). 
We use an iterative procedure. In each iteration we substitute a guess $f_n(z)$ for the tube profile
into the Schr\"odinger equation (\ref{m}). The associated normalized ground state wave function $\phi_n(z)$
and electron ground state energy $\varepsilon_n$ is computed. The wave function $\phi_n$ is then substituted into
Eq. (\ref{o}). This is used as the next guess $f_{n+1}(x)$ for the tube profile, and the process is repeated until
the change in electron energy $|\varepsilon_{n+1}-\varepsilon_{n}|$ is smaller than the required accuracy $\delta$.
We choose $\delta=10^{-4}$, and an initial guess for the tube profile
\begin{equation}
f_1(z)=-\frac{1}{6} z^2(3-z).\label{v}
\end{equation}
This would have been the exact tube profile (cf. Eq.~(\ref{o})), had the electron been localized right at the 
tip (z=1) of the tube. 
In each iteration the total energy is also calculated, and we check that it decreases in each
iteration of the calculation. This gaurantees that the obtaind solution is a minimum.

We find that for $\alpha$ larger than about $1.1\alpha_{\rm min}$, convergence is obtained
within less than $20$ iterations, while for smaller $\alpha$ up to $120$ iterations are required. 
In Figure \ref{f3} we show (converged) tube profiles and wave functions calculated with
this procedure for three different values of $\alpha$.

The value of $\alpha_{\rm min}$, the smallest value of $\alpha$ for which 
polaron states exist, is obtained by repeating the iterative numerical calculation for smaller and 
smaller $\alpha$, until no amount of iteration produces convergence any more. We find that for smaller and
smaller $\alpha$ down to $254.518$, the number of iterations required to obtain convergence slowly increases
up to $120$. Then for $\alpha=254.510$, there is a sudden jump, and after $200$ iterations, convergence is still
not obtained. We conclude that $\alpha_{\rm min}=254.51=0.8293\alpha_c$.

\section{The parameter dependence of the phonon-number $n$}
\label{app4}
In Fig. 3 of the main text the phonon-number $n$ is plotted as a function of the coupling constant 
$\alpha$.  
Here we show that indeed, $n$ does not depend on $L$ and $r$ separately, but only on $\alpha$, since this
is not clear a priori.
(The above statement holds subject to the approximation $I=\pi g r^3$, which is valid when $g\ll r$.
A more sophisticated approximation for $I$ yields only a very weak $r$ dependence in the
regime of realistic radii.) 

Note firstly that $\omega_0$ is given by
\begin{equation}
\omega_0=\frac{3.52}{L^2}\sqrt{\frac{YI}{\rho}},
\end{equation}                    
where $\rho$ is the mass per unit length of the tube, which is proportional to the tube radius. 
(The proportionality constant is $1.47\times10^4\, m_e/a_0^2$.)
Since $I$ is proportional to $r^3$,  $\omega_0$ is proportional to $r/L^2$. 
The bending energy on the other hand
can be written as
\begin{equation}
U_0=-\frac{\hbar^2}{4m^*L^2}\alpha\int_0^1dz\,f_\phi \phi^2 .
\end{equation}
Since the effective electron mass $m^*$ is proportional to $1/r$, 
$m^*L^2\omega_0$ is independent of $r$ and $L$ so that $n=U_0/\omega_0$ is
a function of $\alpha$ only.

\section{Estimating when $|y(L)|\simeq L$}
\label{app5}
In the main text the phase diagram of the system is discussed. We limit our discussion of the
polaron phase to the regime $|y(L)|<\simeq L$. This gives rise to the upper boundaries of the shaded
regions in Fig.~4 of the main text. The estimate for $y(L)~L$ was obtained as follows:
We firstly note that $|y(L)|<|y_{\rm max}|$ where $|y_{\rm max}|$ is the deflection produced when
the electron is completely localized at $x=L$. From Eq.~\ref{v} we have $|y_{\rm max}|=eEL^3/3IY$. Thus, we demand
that $1>eEL^2/3IY$. 

\acknowledgments
This research was supported by the National Research Foundation (NRF) of South Africa.

\end{document}